# An on-chip III-V-semiconductor-on-silicon laser frequency comb for gas-phase molecular spectroscopy in real-time

Kasper Van Gasse,[1] Zaijun Chen,[2] Edoardo Vicentini,[2,3,4] Jeonghyun Huh,[2] Stijn Poelman,[1] Zhechao Wang,[1] Gunther Roelkens,[1] Theodor W. Hänsch,[2,5] Bart Kuyken,[1*] Nathalie Picqué[2*]

[1]Photonics Research Group, INTEC, Ghent University - imec, Technologiepark-Zwijnaarde 126, 9052 Ghent, Belgium
[2]Max-Planck Institute of Quantum Optics, Hans-Kopfermann-Straße 1, 85748, Garching, Germany
[3] Dipartimento di Fisica, Politecnico di Milano, Piazza L. Da Vinci, 32 20133 Milano, Italy
[4] Istituto di Fotonica e Nanotecnologie, Consiglio Nazionale delle Ricerche, Piazza L. Da Vinci, 32 20133 Milano, Italy
[5]Ludwig-Maximilian University of Munich, Faculty of Physics, Schellingstr. 4/III, 80799, München, Germany
*Corresponding authors: bart.kuyken@intec.ugent.be, nathalie.picque@mpq.mpg.de

**Frequency combs, spectra of evenly-spaced narrow phase-coherent laser lines, have revolutionized precision measurements [1]. On-chip frequency comb generators hold much promise for fully-integrated instruments of time and frequency metrology [2]. While outstanding developments are being reported with Kerr [3], quantum cascade [4] and microring electro-optic [5] combs, the field of high-resolution multiplexed gas-phase spectroscopy [6] has remained inaccessible to such devices, because of their large line spacing, their small number of usable comb lines, the intensity variations between their comb lines, and their limited photonic integrability [7-12]. Here we identify a path to broadband gas-phase spectroscopy on a chip. We design a low-noise III-V-on-silicon comb generator on a photonic chip, that emits a flat-top spectrum of 1400 lines at a repetition frequency of 1.0 GHz, a feature never approached by other ultra-miniaturized comb synthesizers. With dual-comb spectroscopy [6], our near-infrared electrically-pumped laser records high-resolution (1 GHz) sensitive multiplexed spectra with resolved comb lines, in times as short as 5 μs. Isotope-resolved $^{12}C/^{13}C$ detection in carbon monoxide is performed within 1 ms. With further developments, entire high-resolution spectroscopy laboratories-on-a-chip may be manufactured at the wafer scale. In environmental sensing, broad networks of spectrometers could be densely field-deployed to simultaneously monitor, in real time, sources and sinks of greenhouse gases, industrial pollution or noxious emissions of motor vehicles.**

Dual-comb spectroscopy creates new opportunities for gas-phase spectroscopy [6]. A recent interferometric technique based on time-domain interference between two frequency combs of slightly-different line spacing, it measures broad-spectral-bandwidth spectra with a sensitivity granted by laser sources of high brightness. This multiplexed method offers an unparalleled consistency in the spectral measurements, for all spectral elements are simultaneously measured on a single photodetector and therefore are similarly affected by conceivable systematic effects, such as temporal variations in the sample. The single photodetector potentially enables operation in any spectral region. The frequency scale may be calibrated with the accuracy of an atomic clock. The interrogation of the sample by narrow laser lines provides a negligible contribution of the instrumental line shape to the spectral profiles. However, in dual-comb spectroscopy with on-chip comb generators, the solution to high-resolution seems cumbersomely restricted to the tuning of the comb lines for interleaving series of spectra measured at different times [8,12], bringing the shortcomings of scanning spectroscopy. Solutions with off-chip external cavities can decrease the line spacing [13], but the potential for a fully-integrated spectrometer vanishes.

We devise a new class of III-V-semiconductor-on-silicon on-chip comb generators [14,15] and we experimentally identify a scheme, based on one of our devices, that is suited to the demanding requirements of high-resolution molecular spectroscopy (Fig.1a, Supplementary Fig. 1, Methods). An on-chip comb generator of 1-GHz line spacing interrogates a gas sample with narrow molecular transitions. The beam is then optically combined with that of a testing electro-optic comb synthesizer and the two outputs of the combiner interfere onto a balanced photodetector. For maintaining the interferometric coherence between the two comb sources [6], indispensable for artefact-free spectroscopic measurements at high resolution and signal-to-noise ratio, a narrow-linewidth continuous-wave laser is used both for optically injection-locking the III-V-semiconductor-on-silicon laser and for feeding the electro-optic system. Our scheme does not involve any tuning elements or moving parts.

Our comb synthesizer (called hereafter III-V-on-Si comb generator) heterogeneously integrates a III-V gain material (InP/InGaAsP) on a silicon-on-insulator passive photonic chip to produce a mode-locked laser that combines low-loss silicon waveguides and high-quality compound semiconductor materials. The silicon photonic circuitry, which is fabricated in a 200-mm wafer pilot line in a CMOS-fab, forms the cavity and the reflectors of the III-V-on-Si laser. In a subsequent post-processing step, an amplifier and a saturable absorber are patterned using standard III/V processing techniques [16]. By wrapping the 37.4-mm-long silicon cavity in a spiral, the footprint of the device is reduced to 0.7 mm$^2$ (Fig. 1b,c). The electrical power, needed to pump the amplifier and to bias the saturable absorber, is only 140 mW, similar to a battery-operated Kerr comb generator [17]. About 10 hours of operation with a AAA battery would be possible. The light, out-coupled with an optical fiber, exhibits a flat-top comb centered at 187 THz. It spans 1.4 THz in a 10-dB bandwidth (Fig.2a), corresponding to 1397 lines spaced by $f_{rep}$=1.0096 GHz. The repetition frequency $f_{rep}$ of the pulse train has a full-width at half-maximum of 150 Hz at a resolution-bandwidth of 150 Hz, limited by environmental conditions. The contribution of amplified spontaneous emission to the linewidth, which indicates its fundamental limit, is 2 Hz (Fig.2b), two-hundred-fold narrower than the previous best report, for a III-V-on-Si mode-locked laser of 9.4-GHz repetition frequency [18]. Moreover, the two degrees of freedom of the comb, repetition frequency $f_{rep}$ and carrier-envelope offset frequency, can be precisely controlled. By superimposing a radio-frequency sine-wave on the saturable-absorber bias, the repetition frequency can be electrically injection-locked to a reference. Ten microwatts are sufficient (Supplementary

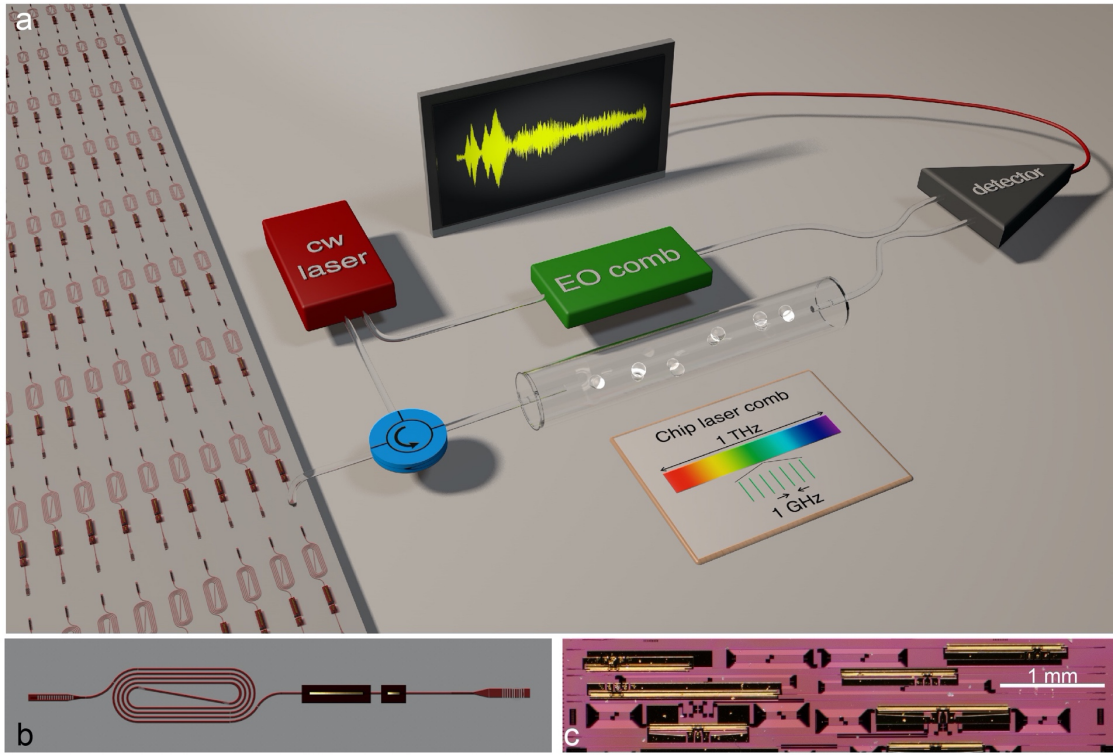

**Figure 1. Gas-phase dual-comb spectroscopy with a III-V-semiconductor-on-silicon comb generator. (a)** Set-up. The light of the III-V-on-Si comb synthesizer is out-coupled from the chip with an optical fiber and interrogates a gas-sample. The beam is multi-heterodyned with that of an electro-optic comb on a balanced photodetector. It contains the interferogram, Fourier transform of the spectrum. For interferometric coherence between the two combs, the continuous-wave laser used for generating the electro-optic comb injection locks one line of the III-V-on-Si comb. **(b)** Schematic representation of the III-V-on-Si laser. **(c)** Photograph of a part of the chip showing 6 mode-locked lasers of different geometries and repetition frequencies.

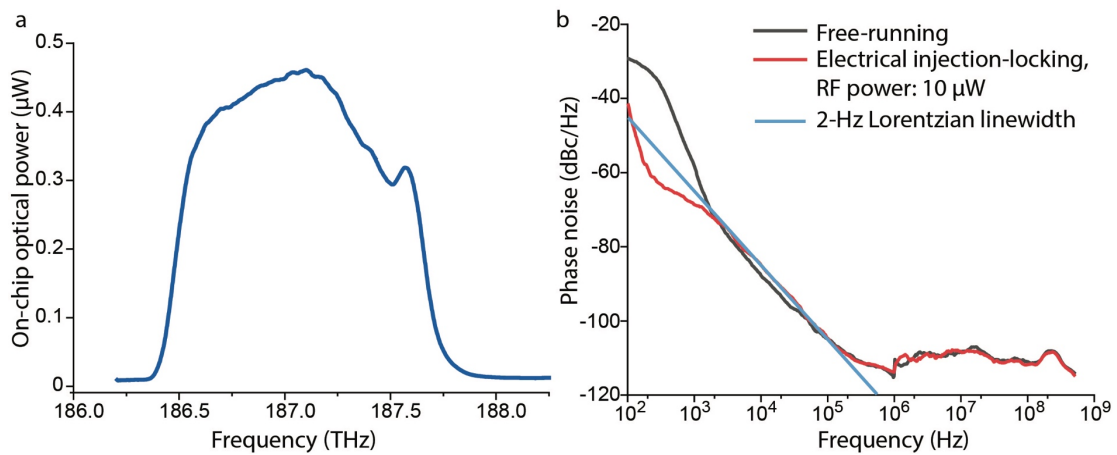

**Figure 2: III-V-on-Si frequency comb generator. (a)** Spectrum, on a linear y-scale, measured with an optical spectrum analyzer. The spectral density of optical power in the comb is scaled to its on-chip value. **(b)** Single-sideband phase noise of the repetition-frequency beat signal in its fundamental mode (1 GHz) in free-running (blue) and electrical injection-locking (red) modes.

Fig.2a), in contrast to quantum cascade lasers which require as much as 6 10$^{-3}$ W [19]. A low-power (on-chip: <10$^{-6}$ W) continuous-wave laser can optically injection-lock the III-V-on-Si comb laser, stabilizing one optical comb line and thus acting on the carrier-envelope offset frequency (Supplementary Fig. 2b).

As we do not possess two matching III-V-on-Si comb generators yet, we interfere one III-V-on-Si comb source with an electro-optic-modulator-based frequency-comb synthesizer, which combines an intensity modulator and a nonlinear fiber [20]. Although its span is limited to 47% of that of the III-V-on-Si comb (Fig.3a), the electro-optic system has a repetition frequency that can be selected by simply dialing a knob and that can match, within any desired offset, the repetition frequency of the III-V-on-Si comb, $f_{rep}$=1.0096 GHz, unlike any other comb generators at our disposal. The two comb beams are combined on a coupler, whose two outputs interfere onto a balanced fast photodetector. The time-domain detector signal is electronically filtered, amplified and digitized to provide the interference pattern. The difference in repetition frequencies of the two combs is $\Delta f_{rep}$=500 kHz. The interferometric signal recurs thus at a period of 2 µs, which sets the minimal time to reach comb-line resolution, although longer measurement times are useful for signal-to-noise-ratio improvement. The coherence time of the interferometer, slightly longer than 100 µs, is limited by the noise generated in the electro-optic system. It improves to 5 ms with a system cascading an intensity and a phase electro-optic modulators (Supplementary Fig. 3), similar to the recent state-of-the-art with on-chip systems [7]. The acquired time-domain sequence is divided in interferograms of 100 µs. The complex Fourier transform of each interferogram reveals amplitude and phase spectra, which are averaged. An averaging time of 120 ms uncovers 660 well-resolved comb lines at high signal-to-noise ratio over a span of 666 GHz (Fig.3b). The number of comb lines is 4 times higher than state-of-the-art dual-comb experiments using on-chip combs [7,11] and our next objective, an on-chip interferometer with two identical III-V-on-Si combs, will straightforwardly increase it two-fold. All across the spectrum, the individual comb lines show the expected instrument signature (Fig.3c,d), with a width that exactly corresponds to the finite measurement time of an interferogram. The signal-to-noise ratio of the comb lines, which culminates at 3.8x10$^4$ for the strongest line, shows a square-root evolution with the averaging time (Fig. 3e).

To demonstrate the potential of the III-V-on-Si comb generator for high resolution gas-phase spectroscopy, we insert a gas sample into its beam path (Fig.1). Molecular absorption in the 187-THz region is weak and we interrogate rovibrational lines (of a full-width at half-maximum of approximately 3 GHz) in the 3-0 band of carbon monoxide, in natural abundance, in a multipass cell of 76-meter path-length. The broad span enables to observe several narrow transitions (Fig.4a) which are properly sampled by the on-chip comb of 1-GHz line spacing (Fig. 4b), contrary to previous reports with microring electro-optic [7] and Kerr [11] combs, where broader profiles were under-sampled by a line spacing of 10 GHz and 22 GHz, respectively. Sampling the profiles at least at the Nyquist condition is necessary to quantitative spectroscopy, such as the unambiguous identification of the absorbents or the retrieval of their concentration. Five microseconds enable a good contrast on a single transition, as shown with a transmittance spectrum of the $P$(21) line in $^{12}C^{16}O$, where only the spectral samples at the comb line positions are plotted (Fig.4c). Owing to the high resolution and the good sensitivity, one millisecond is sufficient to simultaneously detect the $P(21)$ line in $^{12}C^{16}O$ and the $R(11)$ resonance in $^{13}C^{16}O$ (Fig.4d) and to extract an isotopologue ratio of 0.0108(14), within the range of natural abundance (0.0097-0.0117 [21]). Tracing stable isotopologues is important in atmospheric chemistry for monitoring anthropogenic emissions, and in breath analysis for biomarking disease status. To simulate the entire span of the III-V-on-Si comb, we stitch three transmittance (Fig. 4e) and dispersion (Fig. 4f) spectra in the region from the $P(22)$ to the $P(19)$ lines of $^{12}C^{16}O$. Each spectrum, spanning 0.28 THz, results from averaging amplitude and phase spectra over a time of $T$=200 μs. In each spectrum, the average signal-to-noise ratio, defined as the ratio of the signal to the standard-deviation of the fluctuations of the transmittance baseline across the 0.28-THz span, is SNR=56. A metric widely adopted in dual-comb spectroscopy, the figure-of-merit –defined as the product of the average signal-to-noise SNR in unit measurement time multiplied by the number of comb lines $M$(=280)- is $1.1\times10^6$ Hz$^{1/2}$. This is 2.2 and 2,300-fold better, respectively, than that of [11] and that of [7], the only two previous reports of gas-phase multiplexed experiments with combs on photonic chips. The flat-top shape of the envelope of the III-V-on-Si comb contributes to this good figure by providing equally high signal-to-noise ratio over a large span. Our experimental spectra also show very good agreement with spectra computed using the line parameters available in the HITRAN database [22]. The difference between the observed transmittance spectrum and that of HITRAN does not display any systematic signatures.

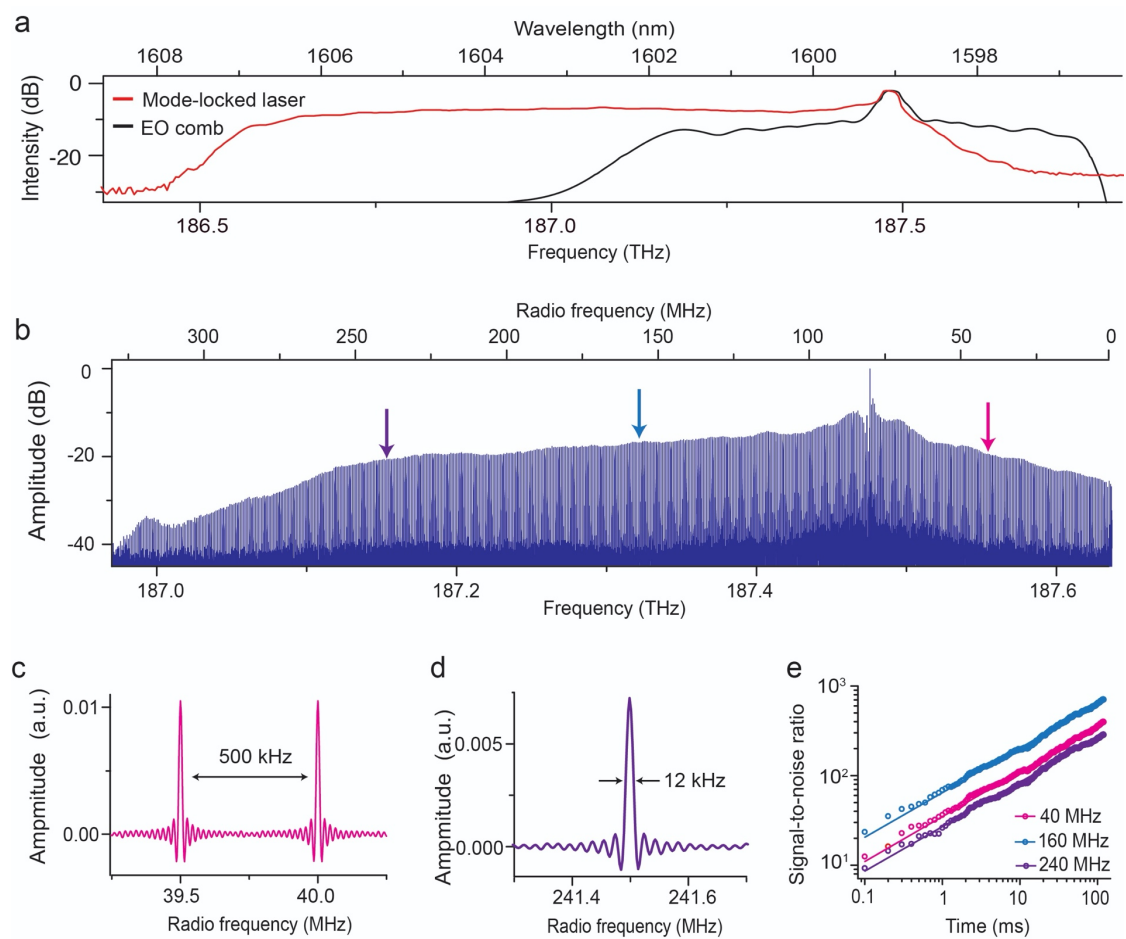

**Figure 3. Experimental dual-comb spectrum resulting from the multiheterodyne beat between the on-chip III-V-on-Si frequency comb generator and an electro-optic-modulator-based comb synthesizer. (a)** Low-resolution optical spectra of the two comb generators. **(b)** Apodized dual-comb spectrum with 660 resolved comb lines and a span limited by the electro-optic-modulator comb, measured within 120 ms. **(c),(d)** Unapodized expanded portions of (b) at different radio-frequencies, showing individual comb lines with the expected instrumental lineshape of transform-limited width **(e)** Evolution of the signal-to-noise ratio of three comb lines across the spectrum with the measurement time: a square-root dependence is observed.

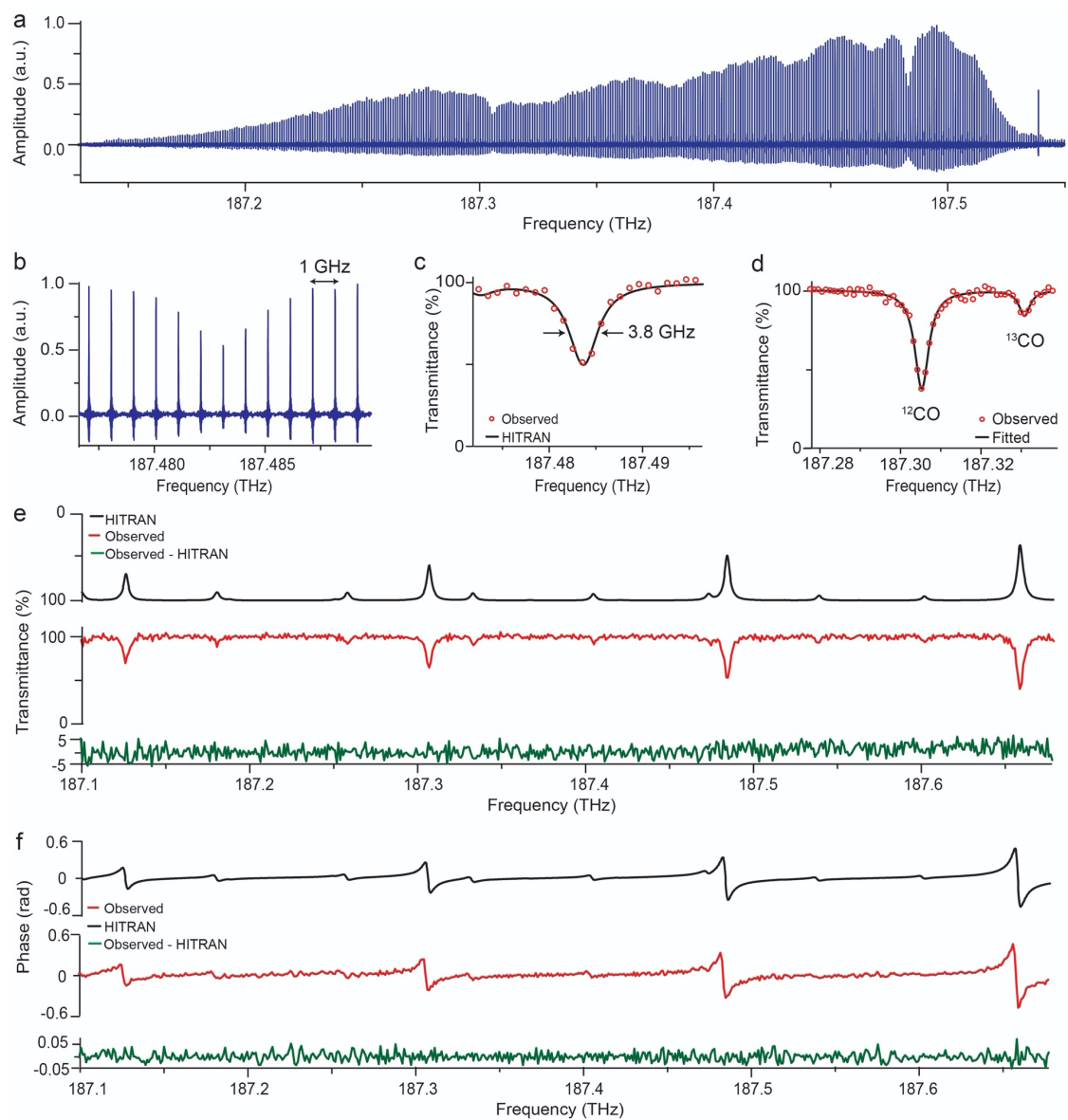

**Figure 4. Experimental comb-line-resolved dual-comb spectra of the 3-0 band in CO at 1-GHz resolution. (a)** Unapodized spectrum with 400 comb lines, spanning 0.4 THz, with the strong *P*(21) and *P*(20) absorption lines of $^{12}C^{16}O$. **(b)** The *P*(20) line sampled by the individual comb lines. **(c)** Spectrum measured within only 5 µs. **(d)** Spectrum measured within 1 ms, showing the *P*(21) line $^{12}CO$ and the *R*(11) line in $^{13}C^{16}O$, for the determination of the $^{13}CO/^{12}CO$ isotopologue ratio. **(e)** Transmittance and **(f)** dispersion experimental spectra in the region of the *P*(22)-*P*(19) lines of $^{12}CO$ and comparison with spectra calculated from the HITRAN database. Three spectra, each spanning 0.28 THz and measured within 200 µs are stitched. The weak lines are assigned to the *R*(9)-*R*(15) lines in $^{13}CO$.

With our proof-of-concept, a significant step is accomplished towards the wafer-scale integration of high-resolution fully-multiplexed gas-phase spectroscopy laboratories-on-a-chip. An on-chip platform for trace-gas detection can be foreseen, where two III-V-on-Si comb generators will be optically injection-locked by a continuous-wave laser and complemented with sample interrogation through sensitive evanescent sensing in a waveguide [23] and fast photo-detection. Moreover, the recent demonstration of epitaxial growth of III-V diodes on silicon [24] and that of heterogeneous integration of interband- and quantum-cascade continuous-wave lasers [25] suggest strategies for III-V-on-Si mode-locked lasers that would directly emit in the mid-infrared molecular fingerprint region, where most molecules have transitions significantly stronger than in the near-infrared.  With improvements to the pulse duration and peak power, self-referencing through spectral broadening in a nonlinear waveguide may be achieved in our conditions of low repetition frequency. Semiconductor-laser combs on a photonic chip might even represent a route towards an on-chip optical-frequency synthesizer for precision measurements [26]. An often-overlooked solution deemed too noisy for actual usefulness, such mode-locked lasers could then be deployed over many applications to ranging, telecommunications, radio-frequency photonics or calibration of astronomical spectrographs.

# Methods

**Fabrication of the III-V-semiconductor-on-silicon comb generator.** The comb generator was fabricated by the die-to-wafer bonding of an active III-V layer stack (containing 6 InGaAsP quantum wells) to a photonic silicon-on-insulator photonic integrated circuit (SOI PIC) chip. The SOI PIC consists of a 750-µm-thick silicon substrate with a 2-µm-thick thermal oxide layer and a 400-nm-thick silicon layer on top. In the 400-nm silicon layer, the waveguides and grating couplers were defined using a 180-nm dry etch into the silicon layer. To attach the III-V material to the SOI PIC a 40-nm thick adhesion layer of divinylsiloxane-bis-benzocyclobutene (DVS-BCB) was used. A gain waveguide was defined in the III-V material after bonding by performing several dry- and wet-etching steps. Finally, the saturable absorber was defined by the electrical isolation of a 40-µm-long and 3-µm-wide section of the gain waveguide, achieved by the dry-etching of the p-contact layer.

**Comb-generator design.** The III-V-on-Si laser was designed using an anti-colliding pulse topology, which ensures that there is only one pulse in the cavity at any time. The cavity is defined by two gratings etched in the silicon waveguide. The saturable absorber is placed close to the low-reflectivity out-coupling grating, while the other grating has a high reflection coefficient. Inside the cavity, the light is coupled from the III-V gain waveguide to the silicon waveguide by the use of an inverted taper (160-µm long) in both the III-V and silicon waveguides. The amplifier has a length of 750 µm, including the inverted taper, and a width of 3 µm. The silicon waveguide has a length of 37.4 mm and is 650-nm wide.

**Comb-generator operation.** The laser is operated on a temperature-controlled chuck and kept at a constant temperature of 290.15 K with an estimated stability of 40 mK. The laser is contacted on-chip with a custom electrical probe containing a low-frequency connection to inject current into the gain waveguide and a high-speed (40-GHz bandwidth) ground-signal-ground path for the saturable absorber. The gain current of 101.05 mA (1.353 V) injected into the amplifier is provided by a DC current/voltage source meter unit. The saturable absorber is biased at -2.895 V generating a photocurrent of -1.7 mA. Mode-locking is self-starting. The operation point is optimized to generate the broadest comb with the lowest pulse-to-pulse timing jitter. The laser has operated in a reproducible way for over 10 weeks (> 1000 operation hours). As calculated from the measured losses of the grating couplers, the laser has an estimated average power of 0.3 mW inside the cavity just before the beam reaches the out-coupling grating. Throughout the manuscript, the on-chip comb power refers to this position. The laser light is extracted using the chip-to-fiber grating coupler. Approximately 15 µW of optical power is coupled to a single-mode fiber, cleaved under 8 degrees to minimize reflections into the laser cavity. If the laser beam were out-coupled on-chip, in a waveguide, the optical power in the waveguide would be on the order of 150 µW. This will be sufficient for an on-chip dual-comb spectrometer, as the power onto the fast detector is usually attenuated to 10-30 µW, to avoid subtle artefacts generated by detector nonlinearities.

The comb generator has a repetition frequency $f_{\text{rep}}$ of approximately 1009600 kHz. The repetition-frequency drift due to a temperature change of the silicon cavity is estimated to 50 kHz·K$^{-1}$. The repetition-frequency beat signal, analyzed by a fast photodiode and a radio-frequency spectrum analyzer, shows a full-width at half maximum of 150 Hz, at a resolution bandwidth of 150 Hz, and a signal-to-noise ratio higher than 60 dB. A single-sideband phase-noise measurement of the repetition-frequency beat signal [18,27,28] (Fig.2d) exhibits, at frequencies below 2 kHz, a flat-top spectrum with a steep edge, corresponding to thermal and acoustic perturbations. Such environmental perturbations can be strongly suppressed by electrical injection locking of the repetition frequency. The second region, from 2 kHz to 400 kHz, shows a $1/f^2$ dependence, signature of the amplified spontaneous-emission noise, where $f$ is the frequency. Finally, the region beyond 500 kHz displays the noise floor of the photoreceiver. The contribution of amplified spontaneous emission to the linewidth of the repetition frequency can be determined by the slope of a $1/f^2$ fit to the phase-noise power density [28]. This contribution is 2 Hz, corresponding to a pulse-to-pulse root-mean-square timing jitter of 17.6 fs.

**Electrical injection-locking of the repetition frequency.** The saturable absorber can be modulated by a signal at a frequency close to the free-running repetition frequency $f_{\text{rep}}$ of the comb on top of the DC bias current, allowing to lock the repetition frequency to a reference clock. A radio-frequency power weaker than 10 µW is sufficient (Supplementary Fig. 2a). The locking cone was determined by sweeping the synthesizer frequency, starting at the free-running frequency. The repetition frequency of the III-V-on-Si comb source follows the synthesizer frequency inside the locking cone.

**Optical injection-locking of a comb line.** One line of the III-V-on-Si comb can be locked to a continuous-wave single-frequency laser by optical injection-locking. Roughly speaking, this equates controlling the comb carrier frequency. The tunable continuous-wave laser emits in the range 184-198 THz (1510 – 1630 nm) with a free-running linewidth of 10 kHz (at 5 µs). A polarization controller and a power attenuator, on its beam path, enable to control the locking conditions. An on-chip optical injection-locking power of 2.5 10$^{-7}$ W is enough, defined as the incoming power in the comb laser cavity, right after the grating coupler. A circulator allows to inject on-chip the continuous-wave-laser beam and to extract the III-V-on-Si-comb light. As discussed below, this scheme also proves instrumental for establishing coherence between the III-V-on-Si comb and the testing electro-optic comb sources. To determine the optical locking range (Supplementary Fig. 2b), the frequency of the continuous-wave laser is swept in steps of 1 MHz. The dual-comb interferogram, visualized in real-time, enables to monitor the locking. Inside the locking cone, the dual-comb spectrum is as shown in Fig. 3b. Outside the locking cone, the signal-to-noise ratio of the beat notes collapses.

**Electro-optic comb generators.** We employ a testing comb generator that cascades an intensity electro-optic-modulator and a nonlinear fiber. The generator is a refined version of that described in [20]. The continuous-wave laser described above also feeds an intensity Mach-Zehnder modulator, of a bandwidth of 20 GHz and an extinction ratio of 53 dB. It is driven by a 50-ps electrical pulse generator of variable repetition frequency set by a low-noise synthesizer. Here the repetition frequency is $f_{rep}+\Delta f_{rep}$ with $\Delta f_{rep}$ ranging from 200 kHz to 1 MHz. The output of the intensity modulator is amplified in a L-band erbium amplifier to an average power of 160 mW. The amplified pulse train is then launched in a nonlinear fiber of a length of 2005 m. The fiber has a dispersion coefficient of -2.57 ps·nm$^{-1}$·km$^{-1}$ at 1550 nm, a dispersion slope of 0.015 ps·nm$^{-2}$·km$^{-1}$ and a non-linear coefficient = 11.7 W$^{-1}$·km$^{-1}$. At the output of the fiber, the spectrum has a span of 775 GHz (at 20-dB bandwidth), leading to 768 comb lines. It is spectrally filtered to optimize the overlap with the III-V-on-Si comb (Fig.3a). The main features of this comb generator are its broad span (for an electro-optic based system) and its flat-top intensity profile. Because it involves a long nonlinear fiber, significant frequency noise is added by minute temperature changes and acoustic perturbations. As discussed below, it has been the limiting factor in the coherence time of the interferometer of about 100 µs (Fig.3b).

Therefore another electro-optic frequency comb generator has been set-up, with the purpose of enabling longer interferometric coherence. An intensity modulator and a phase modulator are cascaded. A sine-wave of frequency 1.008 GHz, sourced by an arbitrary-waveform generator, synchronizes a 50-ps electrical pulse generator, which drives the intensity modulator of a 20-GHz bandwidth. The second driving signal, for the phase modulator, is a sine-wave with a frequency of 16.128 GHz. It feeds a radio-frequency amplifier connected to a phase modulator of 25-GHz bandwidth. The phase between the two sine-waves is empirically adjusted by looking for the broadest and smoothest spectrum on an optical spectrum analyzer. The spectral span is of 201.6 GHz (at 20-dB bandwidth) and comprises 200 comb lines. The spectral enveloped is more modulated than that of the first system, therefore it is not as well suited for molecular spectroscopy. It is only used for the spectrum displayed in Supplementary Fig. 3, which shows a coherence time of 5 ms.

**Dual-comb spectrometer.** At the output of a circulator (Supplementary Figure 1), the fiber-coupled beam of the III-V-on-Si comb generator is amplified to 15 mW in a L-band erbium amplifier. The beam is then collimated and coupled to a compact multipass Herriott-type cell. After 238 passes in the cell, the equivalent absorption path length is increased to 76 m. The mirrors of the cell, not optimized for the 190-THz frequency range, exhibit rather high losses (reflection coefficient of 98.4%), which renders the erbium amplifier necessary. The cell is filled with carbon monoxide in natural isotopic abundance at a pressure of 9.60x10$^4$ Pa. At the output of the cell, the beam is injected in a single-mode fiber. The beam is combined with that of the testing electro-optic modulator comb using a two-input-two-output fibered combiner. The two outputs of the combiner are detected by a balanced photodetector module. A module of a 3-dB bandwidth of 1.6 GHz (respectively 50 MHz) and a noise-equivalent power of 9 pW·Hz$^{-1/2}$ over the 30kHz-100MHz range, (respectively 14.9 pW·Hz$^{-1/2}$) is used in Fig.3,Fig.4a-b (respectively Fig.4c,f,e). The electrical signal contains the interference signal between the two combs. It is electrically low-pass filtered to only collect the signal from one free-spectral range. It is digitized with a data acquisition board (dynamic range: 12 bits, bandwidth: 800 MHz, sampling rate: 1.8 10$^9$ samples·s$^{-1}$).

**Coherence of the dual-comb interferometer by optical injection-locking.** A tunable continuous-wave laser is simultaneously used to optically injection lock the III-V-on-Si mode-locked laser and to feed the electro-optic system. The use of the same laser for both tasks is valuable for maintaining mutual coherence in the dual-comb interferometer. The downside of this scheme is that the radio-frequency spectrum is folded about the continuous-wave laser frequency $f_{cw}$: the frequency $f_{cw}$ is mapped at zero frequency in the radio-frequency dual-comb spectrum and the beat notes between the pairs of comb lines symmetric about the continuous-wave laser line would overlap. To slightly expand the spectral span of the dual-comb spectrum, the beam that is employed for injection locking is frequency-shifted by 80 MHz (Fig. 3) or 25 MHz (Fig.4a,b) using an acousto-optic frequency shifter. This frequency-shifting technical trick is commonly used in electro-optic-modulator-based dual-comb set-ups [7,20,29]. In the conditions of our experiment, this does not prove sufficient to heterodyne the entire III-V-on-Si comb and spectral filtering of the electro-optic modulator comb prevents further aliasing.

The repetition frequencies of the two combs are controlled by synthesizers, which are synchronized, like all electronics in the set-up, to a radio-frequency clock.

In these conditions, an interferometric coherence of 5 ms is experimentally achieved (Supplementary Fig. 3).

**Computation of the spectra.** The time sequences are sliced in interferograms of 100 µs, shorter than the coherence time of the interferometer. Four times as many zeros as there are sample points are added to each side of the interferograms, providing interpolation in the spectrum. When the difference in repetition frequencies is $\Delta f_{rep}$=500 kHz (Fig.3), the zero-optical delay in the interferograms of 100-µs recur fifty times, which allows for well-resolved comb lines. The interferograms are transformed using a complex Fourier transform. Their amplitude and phase spectra are averaged.

**Description of the spectra.** The optical spectrum of the injection-locked III-V-on-Si comb generator and that of the testing electro-optic comb synthesizer only partly overlap, for the injection-locking line at 187.48 THz is shifted from the center of the electro-optic comb by only 80 MHz (Fig.3a). The repetition frequency of the III-V-on-Si comb is $f_{rep}$=1.0096 GHz and that of the testing electro-optic system of $f_{rep}+\Delta f_{rep}$ =1.0101 GHz. An apodized amplitude dual-comb spectrum, recorded without absorbing sample, shows more than 660 resolved comb lines spaced in the radio-frequency domain by $\Delta f_{rep}$=500 kHz. The spectrum is shown on its radio-frequency x-scale and spans 330 MHz, corresponding in the optical domain to 666 GHz. Its measurement time of 0.120 s results of 1200 averaged spectra, each measured within 100 µs. The intense beat note at 80 MHz corresponds to the laser line used for injection locking the III-V-on-Si laser. Its signal-to-noise ratio is $3.8 \cdot 10^4$. Outside this line, the strongest line in the spectrum at 79 MHz has a signal-to-noise ratio of $8 \cdot 10^3$. The average signal-to-noise for the 660 observed comb lines is 800, corresponding to 2310 $Hz^{1/2}$ for a unit measurement time. A portion of the unapodized spectrum at a radio-frequency around 40 MHz shows two individual lines (Fig. 3c), spaced by $\Delta f_{rep}$=500 kHz, while another one displays a comb line at the other end of the spectrum, at 241.5 MHz (Fig.3d). The lines exhibit the expected profile, a cardinal sine which is the instrumental line shape of the spectrometer: the interferogram is measured over a finite time window, it is therefore multiplied by a boxcar signal of a width of the time span. The Fourier transform of the interferogram– the spectrum – is convolved by a cardinal sine whose full-width at half-maximum is 1.2 times the inverse of the time span of the transformed sequence. The observed full-width at half-maximum of the comb lines is 12 kHz, corresponding exactly to the transform limit of a time span of 100 µs. All across the spectrum, the signal-to-noise ratio of the comb lines evolves with the square-root of the averaging time, as illustrated for three comb lines located at 40, 160 and 240 MHz, respectively: on a logarithmic x- and y-scales, the signal-to-noise ratio increases linearly with a slope of 0.5 (Fig.3e).

With a 76-m multipass cell filled with carbon monoxide in natural abundance at a pressure of $9.60 \times 10^4$ Pa and a temperature of 296 K, an unapodized spectrum reveals the imprint of the $P(21)$ and $P(20)$ lines of the 3-0 band of $^{12}CO$ onto the comb lines (Fig. 4a). The difference in repetition frequencies between the two combs is $\Delta f_{rep}$ = 1 MHz. Twenty spectra, each acquired within 100 µs, are averaged for a total measurement time of 2 ms. The amplitude spectrum spans 0.4 THz with 400 resolved comb lines. An expanded portion of this spectrum magnifies the P(20) line of the 3-0 band in $^{12}CO$, well sampled by the individual comb lines of the III-V-on-Si comb with a spacing of 1.0096 GHz. The collisional full-width for self-broadening is 3.26 GHz and 3.09 GHz, respectively, for the $P(19)$ and $P(20)$ lines [30]. As the Doppler full-width is 433 MHz, the molecular profile can reasonably be described by a Lorentzian function. In the transmittance spectrum, the expected full-width at half-maximum of 3.8 GHz for the $P(20)$ line is experimentally observed (Fig. 4c).

For a short measurement of a single molecular line, we spectrally filter the two combs to about 50 GHz. A transmittance spectrum, sampled at the position of the 50 comb lines (Fig. 4c), focuses onto the $P(20)$ line, measured within only 5 µs. The signal-to-noise ratio of the transmittance baseline is 45. The observed transition shows good agreement with a spectrum calculated using the line parameters available in the HITRAN database [22]. The standard deviation of the difference "observed – calculated" is 2.2%, at the noise level.

The good signal-to-noise ratio allows for the simultaneous observation of lines of the two most abundant isotopologues, $^{12}C^{16}O$ and $^{13}C^{16}O$, in a sample in natural abundance, within an averaging time as short as $10^{-3}$ s, as illustrated with the neighboring $P(21)$ line of the 3-0 band in $^{12}C^{16}O$ and $R(11)$ line of the 3-0 band in $^{13}C^{16}O$. Unlike in all other experiments in Fig. 4, the spectrum in Fig. 4d is recorded in a configuration where the two comb generators interrogate the sample in the multipass cell, which enhances the sensitivity to weak absorptions. As a consequence, a single photodetector, noisier than the balanced detectors, is employed. The figure-of-merit for the spectrum of Fig.4d is only $2.5 \times 10^5$ Hz$^{1/2}$. Lorentzian profiles fit the two lines. From the retrieved cross-sections and the line intensities $S$ reported in [31] ($P(21)$, 3-0 band, $^{12}C^{16}O$: $S$ = $9.14 \times 10^{-25}$ cm·molecule$^{-1}$. $R(11)$, 3-0 band, $^{13}C^{16}O$. $S$=$1.56 \times 10^{-23}$ cm·molecule$^{-1}$), the isotopologue abundance ratio $\delta(^{13}C^{16}O/^{12}C^{16}O)$ is 0.0108(14). The number in parentheses is the uncertainty in units of the last two digits. The statistical uncertainty of the fitted cross-section of the weak $^{13}C^{16}O$ line amounts 12%, whereas the uncertainty on the line intensities in [31] is 2%. The range for the isotopologue abundance ratio in natural abundance is 0.0097-0.0117 [21].

To illustrate the entire span available from the III-V-on-Si comb, we stich three transmittance and dispersion spectra, sampled at the positions of the comb lines with a spacing of 1 GHz (Fig. 4e,f). A balanced detector module (bandwidth: 50 MHz, noise-equivalent power: 14.9 pW·Hz$^{-1/2}$) detects and subtracts the two outputs of the interferometer. The time-domain interference signal is digitized by a data acquisition board of 65-MHz bandwidth and 16-bit resolution. The difference in repetition frequencies is set to $\Delta f_{rep}$=200 kHz and each radio-frequency spectrum of 280 comb lines is mapped into the range from 0-56 MHz. Each complex (amplitude and phase) spectrum spans 0.28 THz in the optical domain and it is measured within $T$=200 µs. The optical center frequencies of the three spectra are at 187.2 THz, 187.4 THz, and 187.55 THz, respectively. The strong lines in the transmittance and dispersion spectra are assigned to the $P$(22) to $P$(19) lines of the 3-0 band in $^{12}C^{16}O$. The weaker lines, just above the noise level, are assigned to the $R$(9) to $R$(15) lines of the 3-0 band of $^{13}C^{16}O$. For each transmittance spectrum, the average signal-to-noise ratio SNR, calculated as the amplitude of the transmittance baseline divided by the standard deviation of the noise of that baseline over the 0.28-THz span, is 56. The figure-of-merit –defined as the product of the average signal-to-noise SNR in unit measurement time multiplied by the resolution elements $M$ (=280 resolved comb lines), is SNRx$M$x$T^{-1/2}$ =1.1×10$^6$ Hz$^{1/2}$. The figure-of-merit is limited by the detector noise, leaving potential for higher signal-to-noise ratio with a fully-integrated spectrometer. For comparison with the two previous gas-phase multiplexed experiments with comb generators on photonic chips [7,11], we calculate the figure of merit using parameters given in the publications. In [11], $T$=2 10$^{-4}$ s, $M$=173, the standard-deviation between the dual-comb spectrum and a reference spectrum is σ=0.0254, which we interpret as SNR=1/σ=40, leading to SNRx$M$x$T^{-1/2}$= 4.9 10$^5$ Hz$^{1/2}$. In [7], $T$=95 s, $M$=160, the standard-deviation between the dual-comb spectrum and a reference spectrum is σ=0.034, which we interpret as SNR=1/σ=29, leading to SNRx$M$x$T^{-1/2}$= 4.8 10$^2$ Hz$^{1/2}$.

In our experiment, the noise-equivalent absorption coefficient in unit measurement time per comb line, defined as $(L_{abs}$ SNR$)^{-1}(T/M)^{1/2}$, is 4.0 × 10$^{-9}$ cm$^{-1}$·Hz$^{-1/2}$, where $L_{abs}$(38 m =76/2 m) is the equivalent absorption path length, which takes into account that only one comb beam interacts with the sample. The transmittance and dispersion spectra show good agreement with spectra computed using the line parameters available in the HITRAN database. In the transmittance spectrum, the difference between observed and calculated spectra do not show any systematic signatures and its standard deviation is 2%. In the dispersion spectrum, the standard deviation is 0.017 rad.

**Acknowledgments:** We are grateful to Karl Linner for technical support.
**Funding:** European Research Council (ERC) Starting Grant (ELECTRIC). H2020 Marie Curie Innovative Training Network Microcomb (Grant 812818)**.** Carl-Friedrich von Siemens Foundation. Max-Planck Society. Flemish Research Council (FWO) (12ZB520N).

# Supplementary Figures

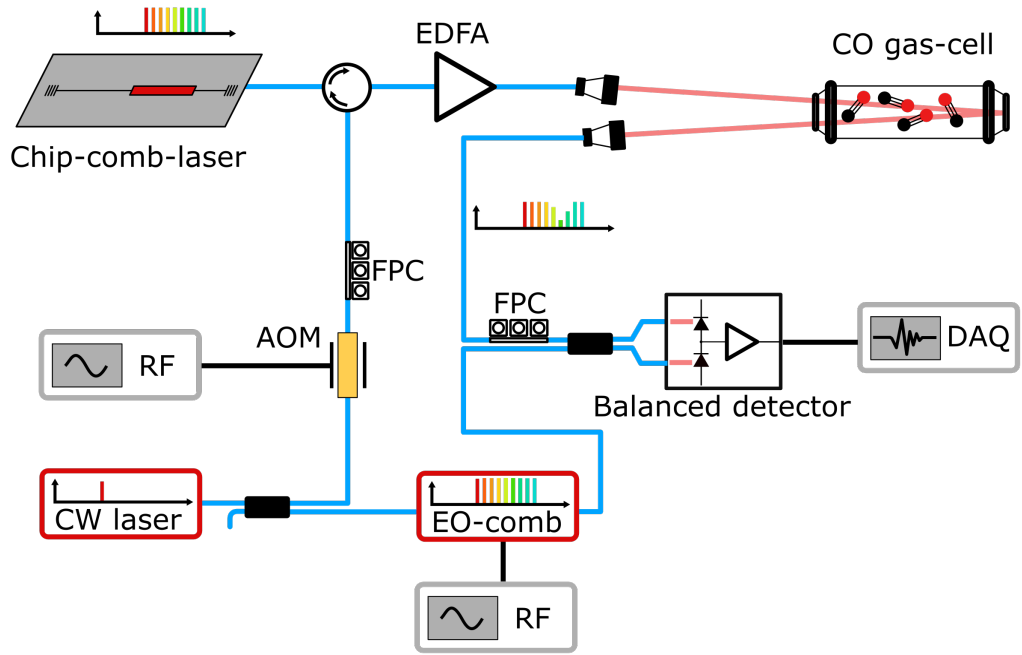

**Supplementary Figure 1: Detailed experimental set-up.** EDFA: Erbium Doped Fiber Amplifier; CO: carbon monoxide; FPC: Fiber Polarisation Controller; DAQ: data acquisition; EO comb: Electro-Optic comb generator; CW laser: Continuous-Wave laser; AOM: Acousto-Optic Modulator.

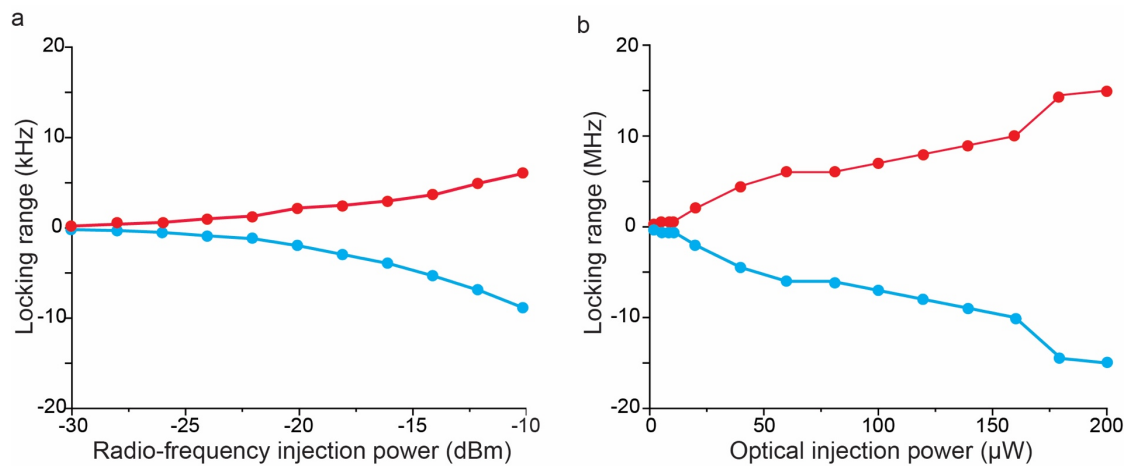

**Supplementary Figure 2: Electrical and optical injection-locking (a)** Locking cone of the repetition frequency with electrical injection locking by modulation of the saturable absorber, down to the µW power level. **(b)** Locking cone with optical injection as a function of the on-chip optical continuous-wave-laser power injected to the photonic chip.

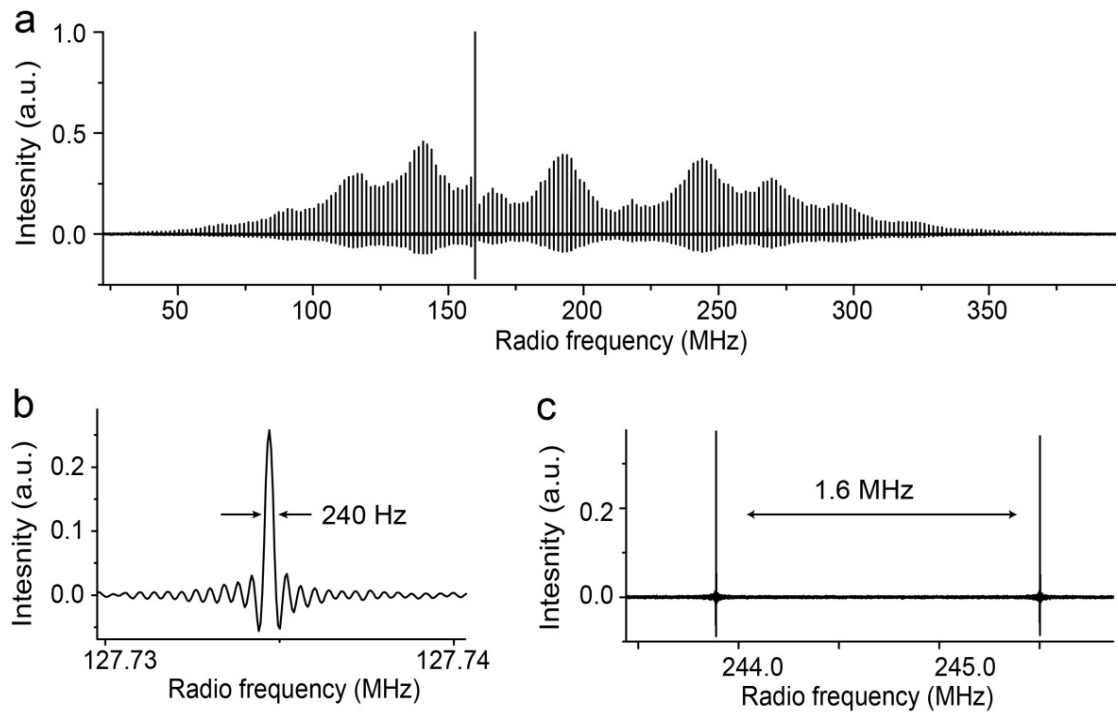

**Supplementary Figure 3: Experimental dual-comb spectrum with an interferometric coherence time of 5 ms.** **(a)** Unapodized spectrum over the radio-frequency span of 200 comb lines, resulting from a Fourier transform of a 5-ms interferogram. The periodic modulation, with a recurrence of 16 comb lines, is due to the phase modulator. The strong line at 160 MHz is the beat note of the continuous-wave laser used for injection-locking, which is shifted by 20 GHz with respect to its replica used for feeding the electro-optic modulators. The signal-to-noise ratio of the comb line at 200 MHz is 520. **(b)** Magnified representation showing a comb line at 127.735 MHz. The full-width at half-maximum of the comb line is 240 Hz, corresponding exactly to the Fourier transform limit. **(c)** Magnified representation showing two comb lines at 244.5 MHz, spaced by $\Delta f_{rep}$=1.6 MHz.